\documentclass[12pt]{article}
\usepackage{graphicx}
\usepackage{hyperref,wrapfig,cite,color,subfigure,amssymb,amsmath,xspace,enumerate}
\usepackage{hyperref}

\newcommand\pubnumber{KA-TP-29-2011\\MCnet-11-25}
\newcommand\pubdate{\today}

\newcommand{\HWPP}{\textsf{Herwig++}\xspace}

\newcommand{\NC}{N_{\mathrm{C}}}

\newcommand{\preco}{\ensuremath{p_{\rm reco}}\xspace}
\newcommand{\pdisrupt}{\ensuremath{p_{\rm disrupt}}\xspace}
\newcommand{\ptlead}{\ensuremath{p_{\perp}^{\rm lead}}\xspace}
\newcommand{\ptsum}{\ensuremath{\sum p_{\perp}}\xspace}

\newcommand{\avgN}{\langle n(\vect b, s) \rangle}
\newcommand{\pt}{\mathmode{p_{\perp}}}
\newcommand{\ptmin}{\mathmode{p_{\perp}^{\rm min}}}

\newcommand{\mathmode}[1]{\relax\ifmmode #1\else{$#1$}\fi}

\newcommand{\vect}[1]{{\bf #1}}

\newcommand{\seff}{\mathmode{\sigma_{\rm eff}}}

\newcommand{\sigmahard}{\mathmode{\sigma^{\rm inc}_{\rm hard}}}
\newcommand{\dNchgdetadphi}{\ensuremath {\langle \mathrm{d}^2N_\text{ch}/\mathrm{d}\eta\,\mathrm{d}\phi\rangle} \xspace}
\newcommand{\dpTsumdetadphi}{\ensuremath{\langle \mathrm{d}^2\!\sum\!\pt/\mathrm{d}\eta\,\mathrm{d}\phi \rangle}\xspace}

\def\napoli{Institut f\"ur Theoretische Physik, Karlsruhe Institute of Technology, Karlsruhe,
Germany}

\def\Title#1{\begin{center} {\Large #1 } \end{center}}
\def\Author#1{\begin{center}{ \sc #1} \end{center}}
\def\Address#1{\begin{center}{ \it #1} \end{center}}

\newcommand\pubblock{\rightline{\begin{tabular}{l} \pubnumber\\
         \pubdate  \end{tabular}}}
\newenvironment{Abstract}{\begin{quotation}  }{\end{quotation}}
\newenvironment{Presented}{\begin{quotation} \begin{center} 
             PRESENTED AT\end{center}\bigskip 
      \begin{center}\begin{large}}{\end{large}\end{center} \end{quotation}}

\def\beq{\begin{equation}}
\def\eeq#1{\label{#1}\end{equation}}
\def\eeqn{\end{equation}}

\def\beqa{\begin{eqnarray}}
\def\eeqa#1{\label{#1}\end{eqnarray}}
\def\eeqan{\end{eqnarray}}

\let\bar=\overbar

\def\Dslash{\not{\hbox{\kern-4pt $D$}}}
\def\dslash{\not{\hbox{\kern-2pt $\del$}}}

\def\msb{{\bar{\ssstyle M \kern -1pt S}}}

\begin{document}
\begin{titlepage}
\pubblock

\vfill
\Title{Multiple Partonic Interaction\\
Developments in Herwig++
}
\vfill
\Author{ S.~Gieseke, C.~A.~R\"ohr, A.~Si\'odmok\footnote{Speaker.}
}
\Address{\napoli}
\vfill
\begin{Abstract}
We briefly review the status of the multiple partonic interaction model in the \HWPP event
generator. First, we show how a change in the colour structure of an event in \HWPP
results in a significant improvement in the description of soft inclusive observables in
$pp$ interactions at $\sqrt{s}=900$ GeV. Then we present a comparison of some model
results to ATLAS Underlying Event data at $\sqrt{s}=7$ TeV.
\end{Abstract}
\vfill
\begin{Presented}
MPI@LHC 2010: 2nd International Workshop on Multiple Partonic Interactions at the LHC\\
Glasgow, 29th of November to the 3rd of December 2010
\end{Presented}
\vfill
\end{titlepage}
\def\thefootnote{\fnsymbol{footnote}}
\setcounter{footnote}{0}

\section{Introduction}
The magnificent operation of the LHC in 2010 gave us the opportunity to study physics
at the new high-energy frontier. The first physics results from the LHC experiments were
measurements of Minimum-Bias (MB) \cite{:2010ir}  and Underlying Event (UE)
characteristics \cite{Aad:2010fh}. Understanding the UE and  MB interactions is very
important for many aspects of LHC physics. The amount of UE activity at the LHC is
measured, so one might think that the size of the UE correction is known. However, in
practice, there are observables that depend on correlations or fluctuations away from
average properties of the UE, including, to varying extents, any measurement relying on
jets or isolation criteria. In fact, almost every observable that will be used for new
physics searches or precision measurements falls into this class, so the correction must
be represented by a model tuned to data, rather than by a single number measured from
data.  In this short note we present recent developments in the modelling of the multiple
partonic interactions in \HWPP and show, for the first time, a comparison of the
improved model to  $7$ TeV UE data.

\section{Multiple Parton Interactions in \HWPP}
\subsection{Eikonal model}
The first model for hard multiple partonic interactions in \HWPP was implemented 
in version 2.1 of the program and is based on the eikonal model described in 
Ref.~\cite{Butterworth:1996zw}. This model derives from the assumption that at 
fixed impact parameter, $b$, individual scatterings are independent and that the
distribution of partons in hadrons factorizes with respect to the $b$
and $x$ dependence.  This implies the average number of partonic collisions at a given $b$
value to be
\begin{equation}\label{eqn:avgN}
  \avgN = A(b;\mu) \ \sigmahard(s;\ptmin)\, ,
\end{equation}
where $A(b;\mu)$ is the partonic overlap function of the colliding hadrons
and \sigmahard\ is the inclusive production cross section of a
pair of partons with $\pt > \pt^{\rm min}$.
The impact parameter dependence of partons in a hadron is modelled by the
electromagnetic form factor,
\begin{equation}\label{eqn:overlap}
    A(b;\mu) = \frac{\mu^2}{96 \pi} (\mu b)^3 K_3(\mu b) \, ,
\end{equation}
where $K_3(x)$ is the modified Bessel function of the third kind
and $\mu$ is the inverse proton radius. 
Since the spatial parton distribution is assumed to be similar to the
distribution of charge, but not necessarily identical, $\mu$ is treated as a free parameter. 
This model allows for the simulation of multiple interactions with perturbative scatters
with $p_{\perp} > \ptmin$. Due to the lack of soft scatters below $\ptmin$, this first model is
only able to describe the jet production part of the CDF data \cite{Affolder:2001xt},
above approximately 20 GeV, but not the more inclusive minimum-bias part. 

An extension of the model to include soft scatters (with $p_{\perp} < \ptmin$) has been
implemented in \HWPP and has been the default underlying event model as of version 2.3.
The additional soft contribution to the inclusive cross section is also eikonalized. In
this way we can calculate the average number of soft scatters from the respective soft
inclusive cross section $\sigma^{\rm inc}_{\rm soft}$ and the overlap function for the
soft scattering centres $A(b;\mu)$. The functional form of $ A(b;\mu)$ is assumed to be
the same as for the hard scatters, but we allow for a different inverse radius, $\mu_{\rm
soft}$.  We keep this model consistent with unitarity by fixing the two additional
parameters $\sigma^{\rm inc}_{\rm soft}$ and $\mu_{\rm soft}$ from two constraints. First,
we can calculate the total cross section from the eikonal model and fix it to be
consistent with the Donnachie--Landshoff (DL) parametrization \cite{Donnachie:1992ny}.  In
addition, using the optical theorem, we can calculate the elastic $t$-slope parameter from
the eikonal model and fix it to a reasonable parametrization.  This model is
capable of describing the whole spectrum of UE data from the Tevatron including its
minimum bias part. 
  
\subsection{Colour correlations}
\begin{figure}[t]
  \begin{center}
    \subfigure[]{%
    \label{fig:MB900:a}%
    \includegraphics[width=0.50\textwidth]{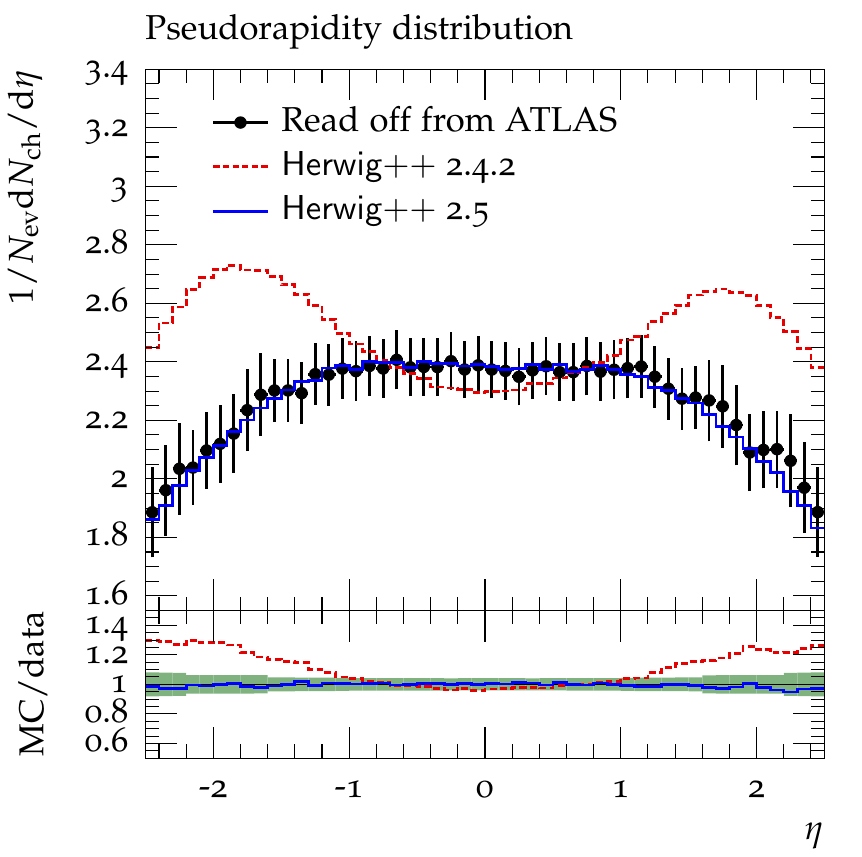}%
    }%
    \subfigure[]{%
    \label{fig:MB900:b}%
    \includegraphics[width=0.50\textwidth]{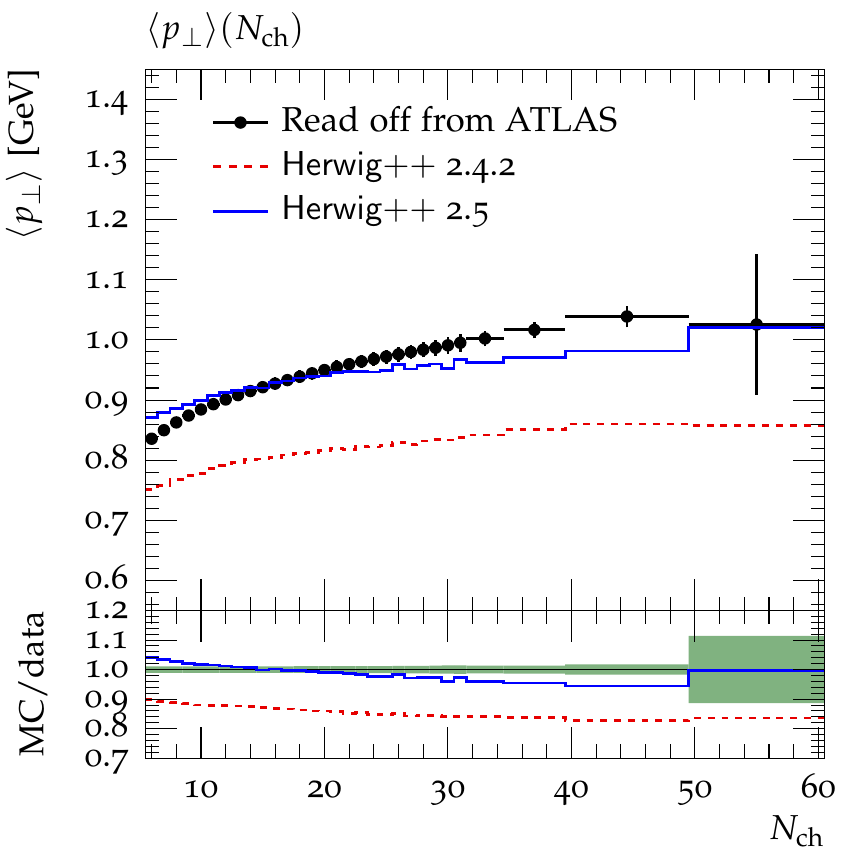}%
    }
    \caption{ Comparison of \HWPP{} 2.4.2 and \HWPP{} 2.5 to ATLAS minimum-bias distributions at
  $\sqrt{s}=0.9~\mathrm{TeV}$ with $N_{\mathrm{ch}} \ge 6$, $p_{\perp} > 500~\mathrm{MeV}$
  and $|\eta| < 2.5$. The ATLAS data was taken from plots published in
  Ref.~\cite{:2010ir}.}%
    \label{fig:MB900}
  \end{center}
\end{figure}

Despite providing a good description of the CDF UE data, this model turned out to be too
simple to describe the Minimum Bias ATLAS data collected at 900 GeV \cite{:2010ir}. In
particular, the predictions for the charged-particle multiplicity as a function of pseudorapidity 
and the average transverse momentum as a function of the particle multiplicity, $\langle p_{\perp}\rangle (N_{\rm
ch})$, are extremely far from the data. This discrepancy is shown in Fig.~\ref{fig:MB900}, where the
red line represents the \HWPP 2.4.2 results, featuring the model as described above\footnote{Currently
 there is no model for soft diffractive physics in \HWPP.  Therefore we
use diffraction-reduced ATLAS MB analysis with an additional cut on the number of charged
particles: $N_{\rm ch} \ge 6$.}. As presented in more detail in \cite{Gieseke:2010zz}, a
tuning of the parameters of the MPI model was not able to improve this description. 

\begin{figure}[t]
  \begin{center}
    \subfigure[]{%
    \label{fig:crsketches:a}%
    \includegraphics[width=0.40\textwidth]{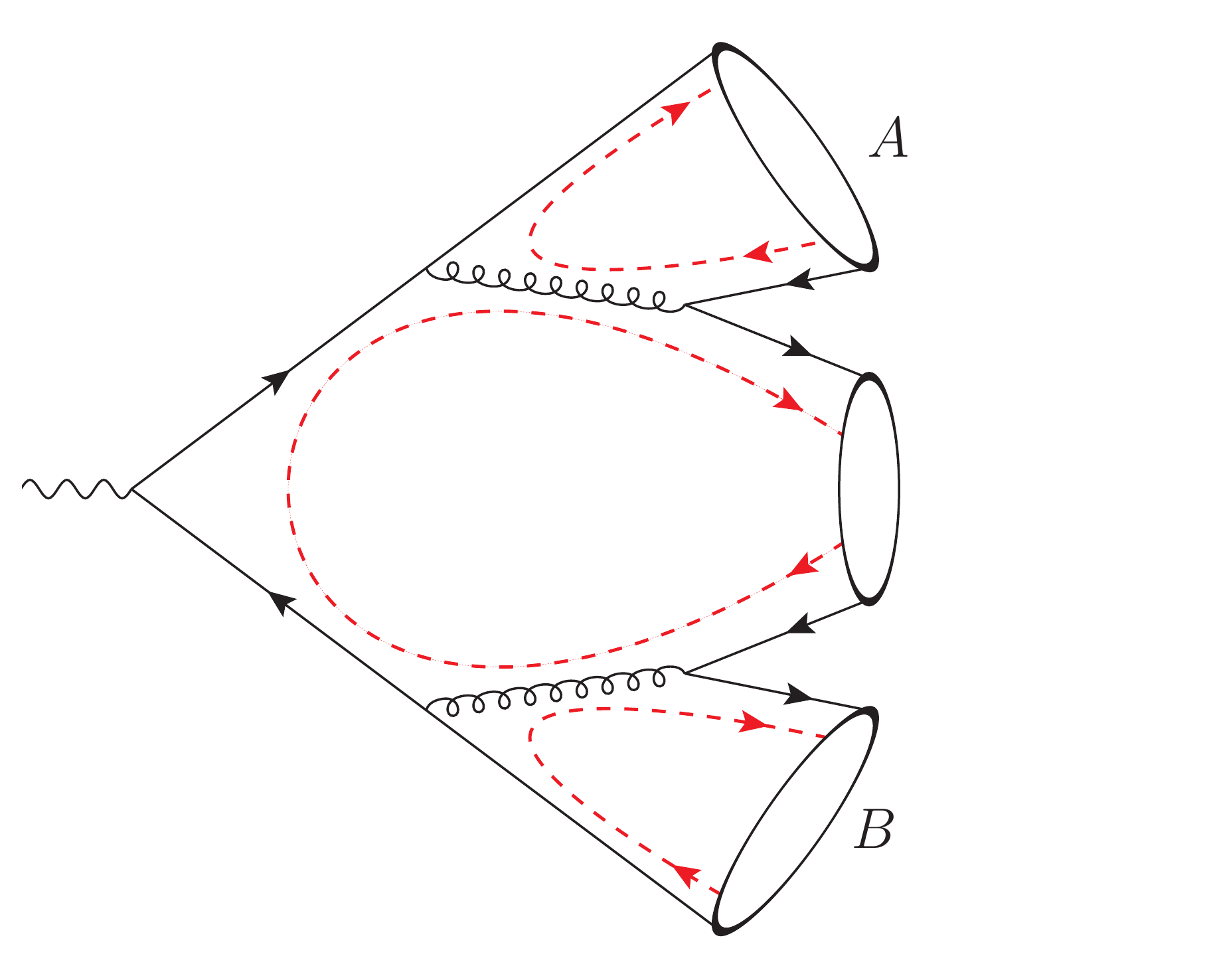}%
    }%
    \subfigure[]{%
    \label{fig:crsketches:b}%
    \includegraphics[width=0.40\textwidth]{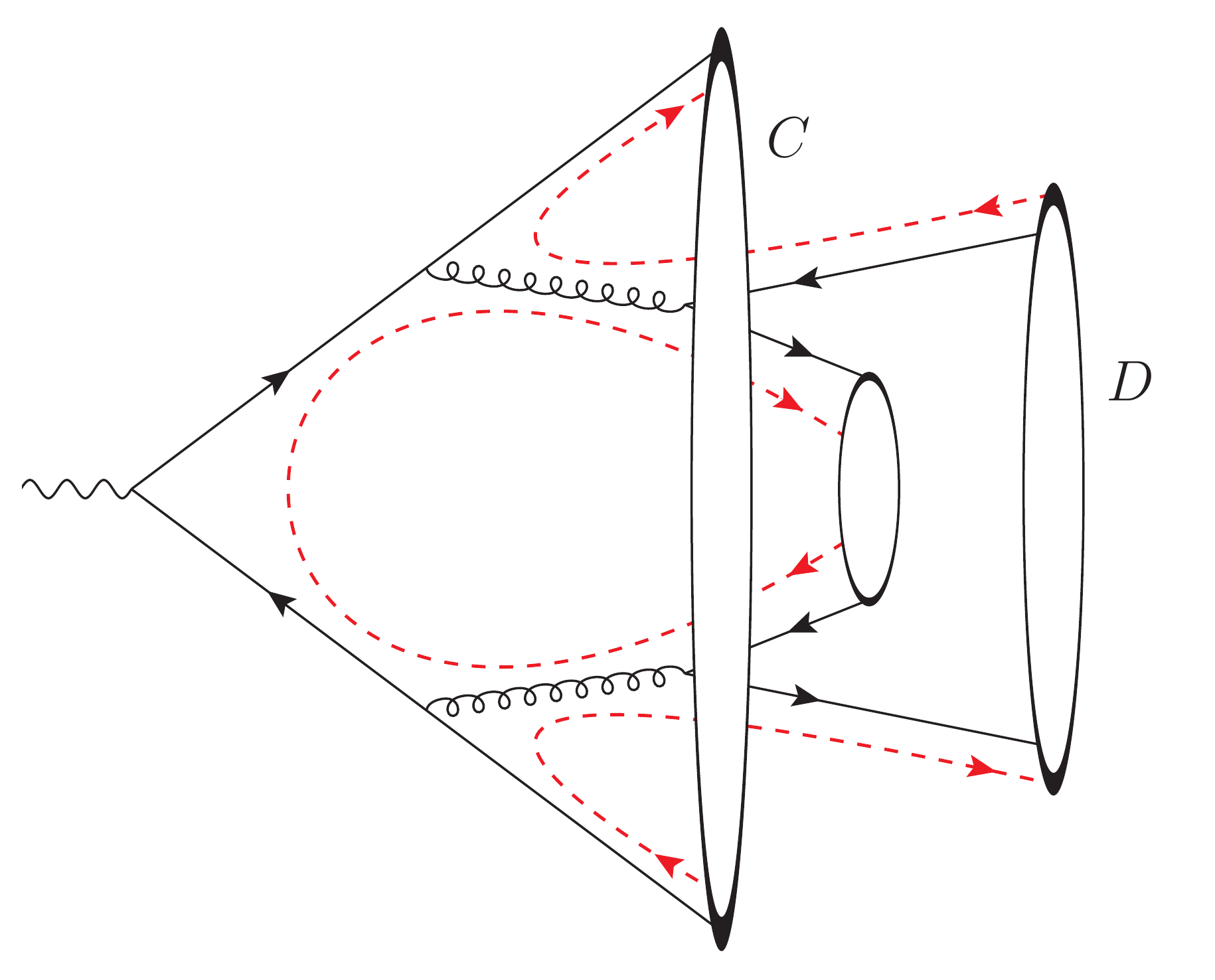}%
    }
    \caption{Formation of clusters, which are represented by ovals. Colour lines are
    dashed. (a) shows colour-singlet clusters formed according to the dominating colour
    structure in the $1/\NC$ expansion. (b) shows a possible colour-reconnected state: the
    partons of the clusters $A$ and $B$ are arranged in new clusters, $C$ and $D$.}%
    \label{fig:crsketches}
  \end{center}
\end{figure}

The prediction of \HWPP for $\langle p_{\perp}\rangle (N_{\rm ch})$ was close to insensitive to
the parameters of the MPI model. Moreover, this observable is known to be very sensitive
to non-perturbative colour reconnection. This triggered new developments of the MPI model to
include non-perturbative colour reconnections (CR). The CR model presented in this
work can be regarded as an extension of the cluster model \cite{Webber:1983if}, which
is used for hadronization in \HWPP~\cite{Bahr:2008pv}.  Hadronization in \HWPP{} is based
on the pre-confinement property of perturbative QCD \cite{Amati:1979fg}. According to
that, a parton shower evolving to the cut-off scale $Q_0$ ends up in a state of colourless
parton combinations with finite mass of $\mathcal{O}(Q_0)$. In the cluster hadronization
model, these parton combinations -- the clusters -- are interpreted as highly excited
pre-hadronic states.  They act as a starting point for the generation of hadrons via
cluster decays, which can be performed in multiple steps. Colour reconnection in the
cluster model occurs at the stage where clusters are formed from the parton-shower
products. Starting with the clusters, produced generically by virtue of pre-confinement,
cf.~Fig.~\ref{fig:crsketches:a}, the cluster creation procedure is slightly modified.
Pairs of clusters are allowed to be `reconnected'. This means the coloured constituent of
cluster $A$ and the anti-coloured constituent of cluster $B$ form a new cluster, as do the
remaining two partons, cf.~Fig.~\ref{fig:crsketches:b}. 
The following steps describe the full algorithm of the colour reconnection model implemented
in \HWPP 2.5:
\begin{enumerate}[1.]
  \item Do the subsequent steps for all quarks in \emph{random} order.
  \item The current quark is part of a cluster. Label this cluster $A$.
  \item \label{clusterfinder} Consider a colour reconnection with any other cluster $B$.
    For the possible new clusters $C$ and $D$, which emerge during reconnecting $A$ and $B$
    as in Fig.~\ref{fig:crsketches}, the following must be satisfied: 
    \begin{itemize}
      \item The new clusters are lighter,
	\begin{equation}
	m_C+m_D < m_A+m_B\ .
	  \label{equ:pcr:condition}
	\end{equation}
	Here, $m_i$ denotes the invariant
	mass of cluster $i$.
      \item $C$ and $D$ do not consist of a $q\bar{q}$ pair produced in a
	preceding gluon splitting.
    \end{itemize}
  \item Amongst the found reconnection possibilities select the one that results in the
    \emph{smallest} sum of cluster masses, $m_C + m_D$. Accept this colour reconnection
    with an adjustable probability \preco. This parameter steers the amount of colour
    reconnection in the CR model.
\end{enumerate}

A further extension of the MPI model is to restore the possibility of changing the colour
connections in the soft component of the model. The model provides a parameter \pdisrupt,
which is the probability of soft scatters to be disconnected in colour space from the rest
of the event.  $\pdisrupt = 1$, the default in \HWPP 2.4, physically corresponds to no
colour strings between the beam remnants and the soft particles produced in the soft
underlying event. 

So in total, there are four main parameters of the MPI model: The inverse hadron
radius squared $\mu$, $\ptmin$, the colour reconnection probability \preco and the probability
of colour disruption of the soft scatters \pdisrupt.

\section{Results}
\begin{figure}[t]
  \begin{center}
    \hfill%
   \subfigure[]{%
   \label{fig:UE7000:a}%
    \includegraphics[width=0.50\textwidth]{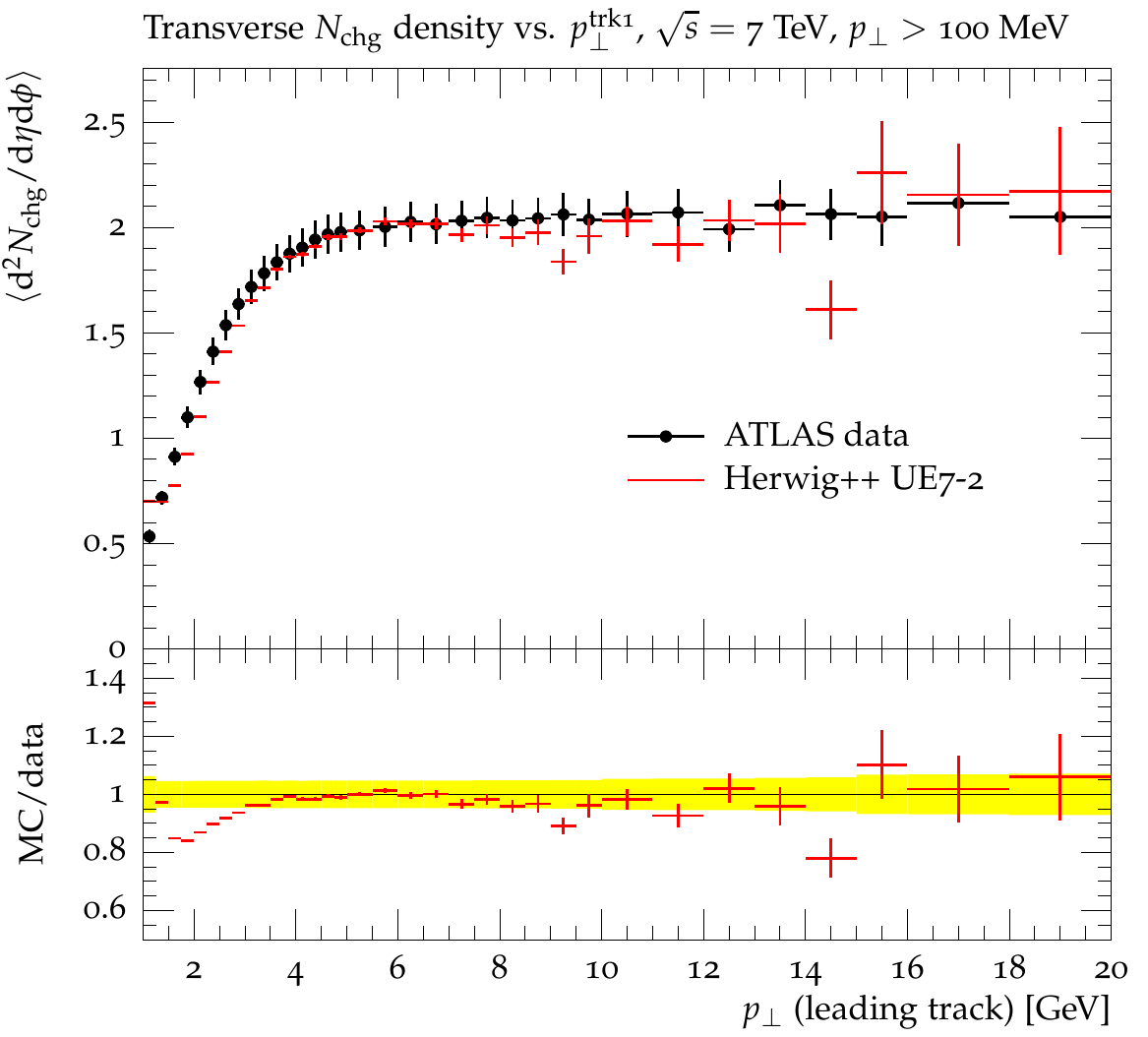}%
   }%
   \subfigure[]{%
   \label{fig:UE7000:b}%
    \includegraphics[width=0.50\textwidth]{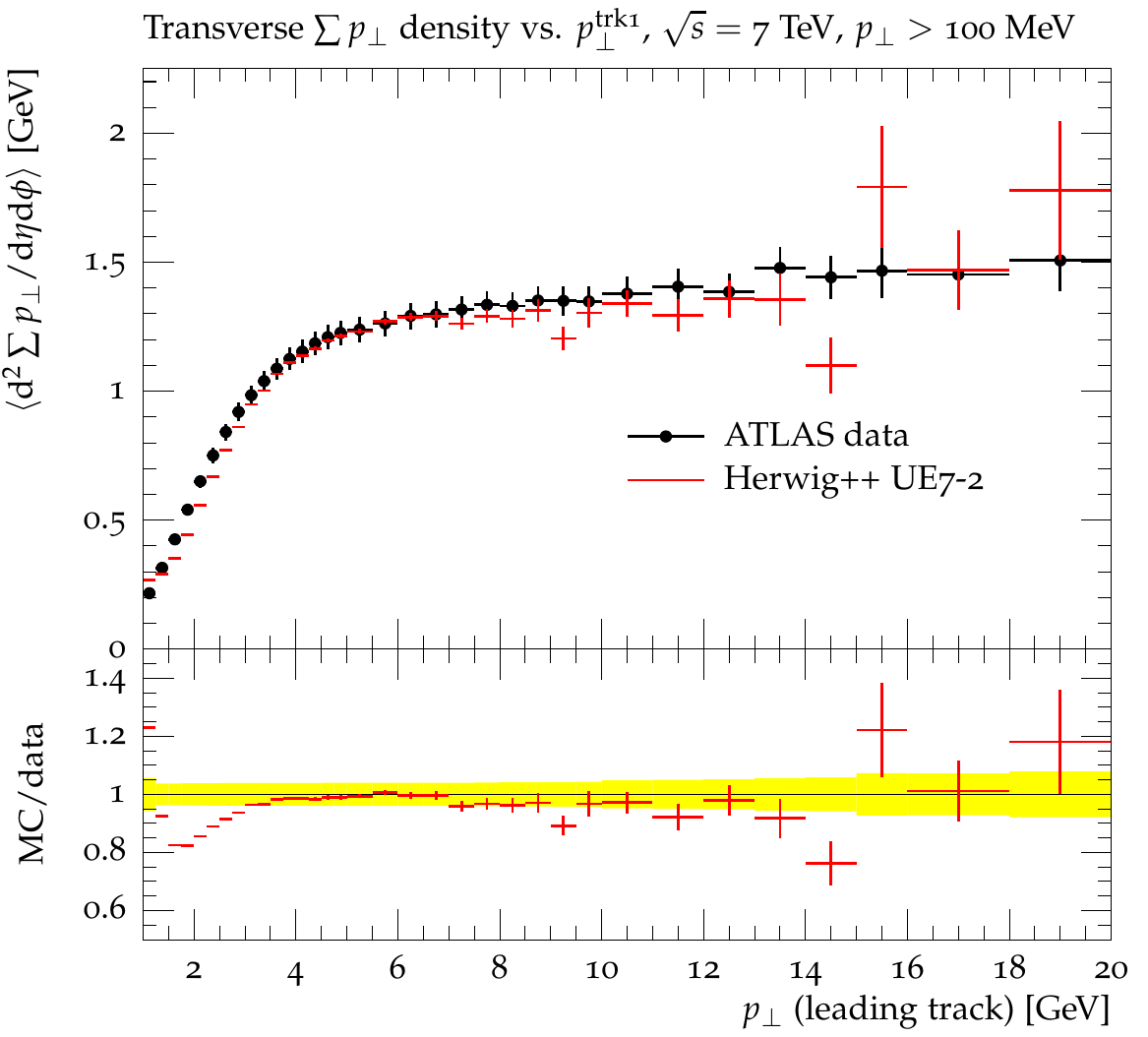}%
}
    \caption{Charged multiplicity and scalar \ptsum density of charged particles with
    $p_t > 100$~MeV and $|\eta| < 2.5$ in the transverse region. The predictions of the
    UE7-2 tune are compared to ATLAS UE data at 7 TeV \cite{Aad:2010fh}.}%
    \label{fig:UE7000}
  \end{center}
\end{figure}

\begin{figure}[t]
  \begin{center}
    \hfill%
    \subfigure[]{%
    \label{fig:UE7000sd:a}%
    \includegraphics[width=0.50\textwidth]{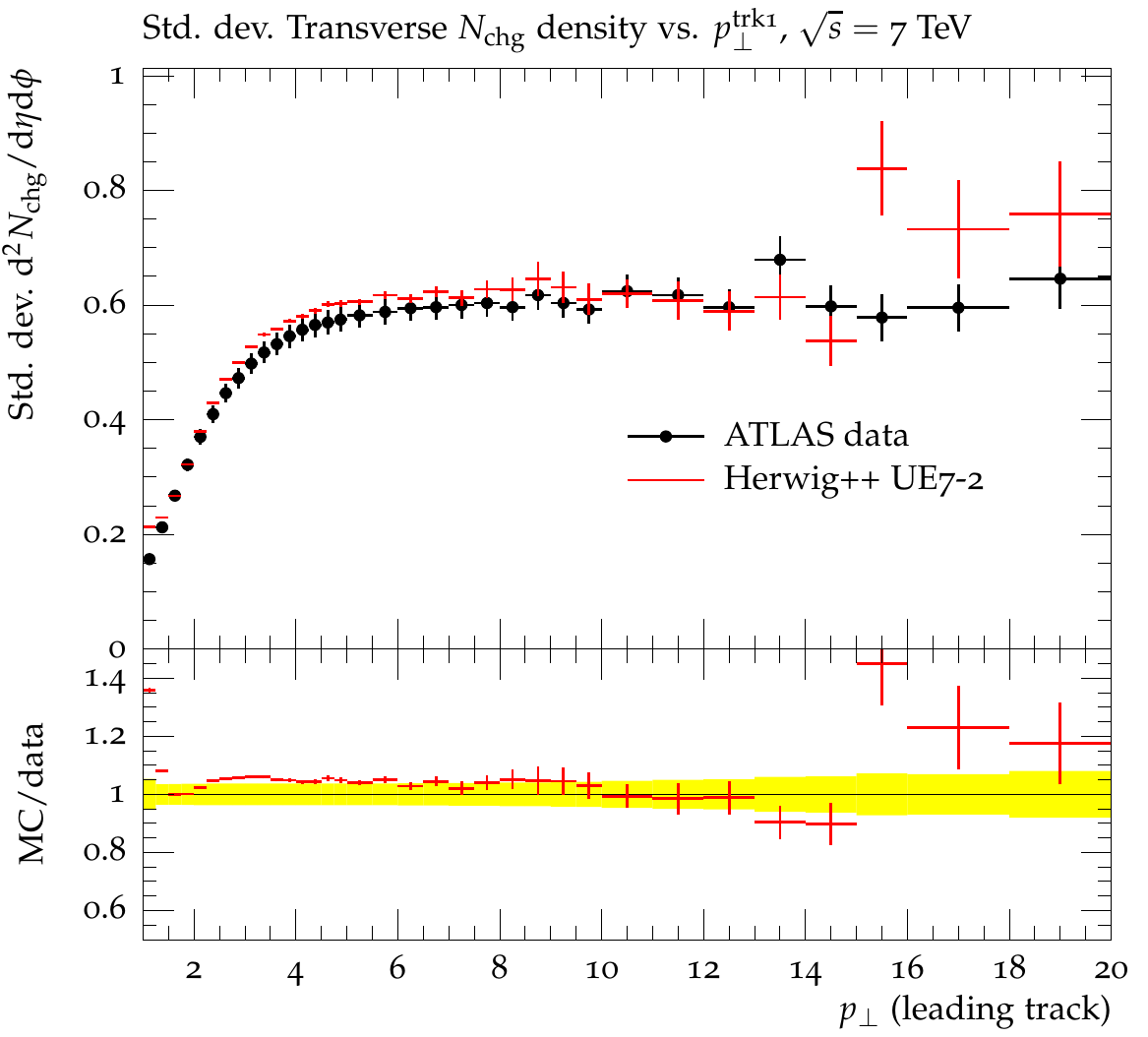}%
    }%
    \subfigure[]{%
    \label{fig:UE7000sd:b}%
    \includegraphics[width=0.50\textwidth]{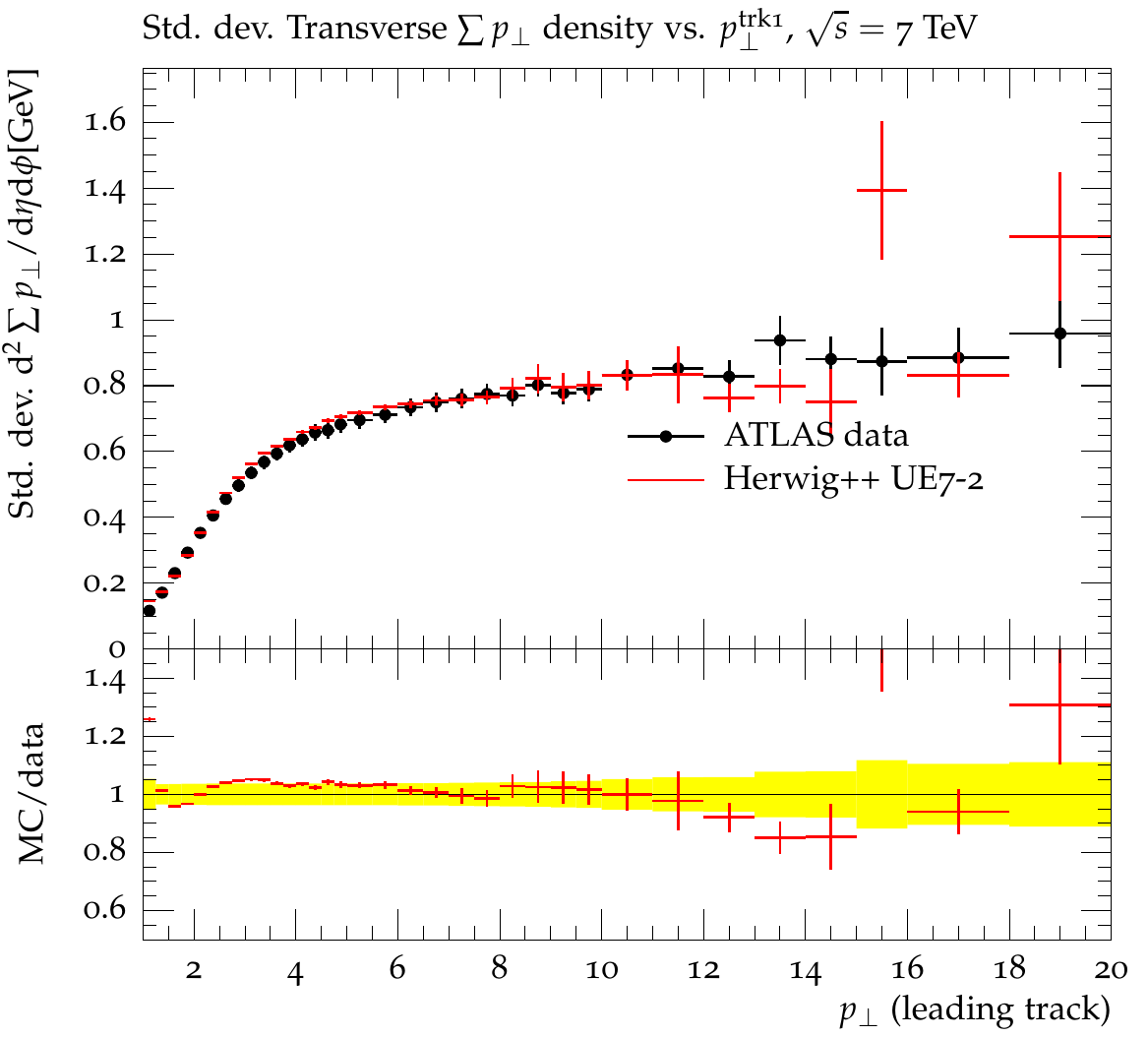}%
    }
    \caption{Standard deviations of the charged multiplicity density and scalar \ptsum
    density of charged particles with $p_t > 100$~MeV and $|\eta| < 2.5$ in the transverse
    region. The predictions of the UE7-2 tunes are compared to ATLAS UE data at
    7 TeV \cite{Aad:2010fh}.}%
    \label{fig:UE7000sd}
  \end{center}
\end{figure}

\begin{figure}[t]
  \begin{center}
    \hfill%
    \subfigure[]{%
    \label{fig:UE7000phi:a}%
    \includegraphics[width=0.50\textwidth]{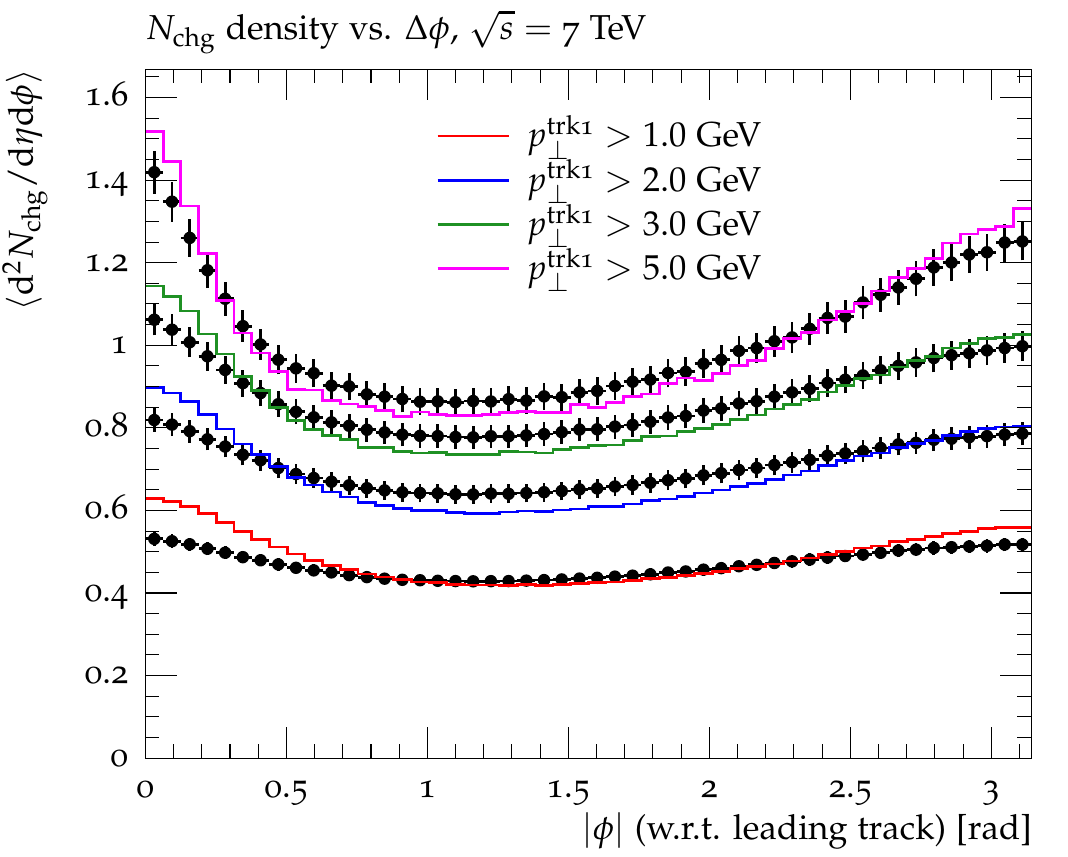}%
    }%
    \subfigure[]{%
    \label{fig:UE7000phi:b}%
    \includegraphics[width=0.50\textwidth]{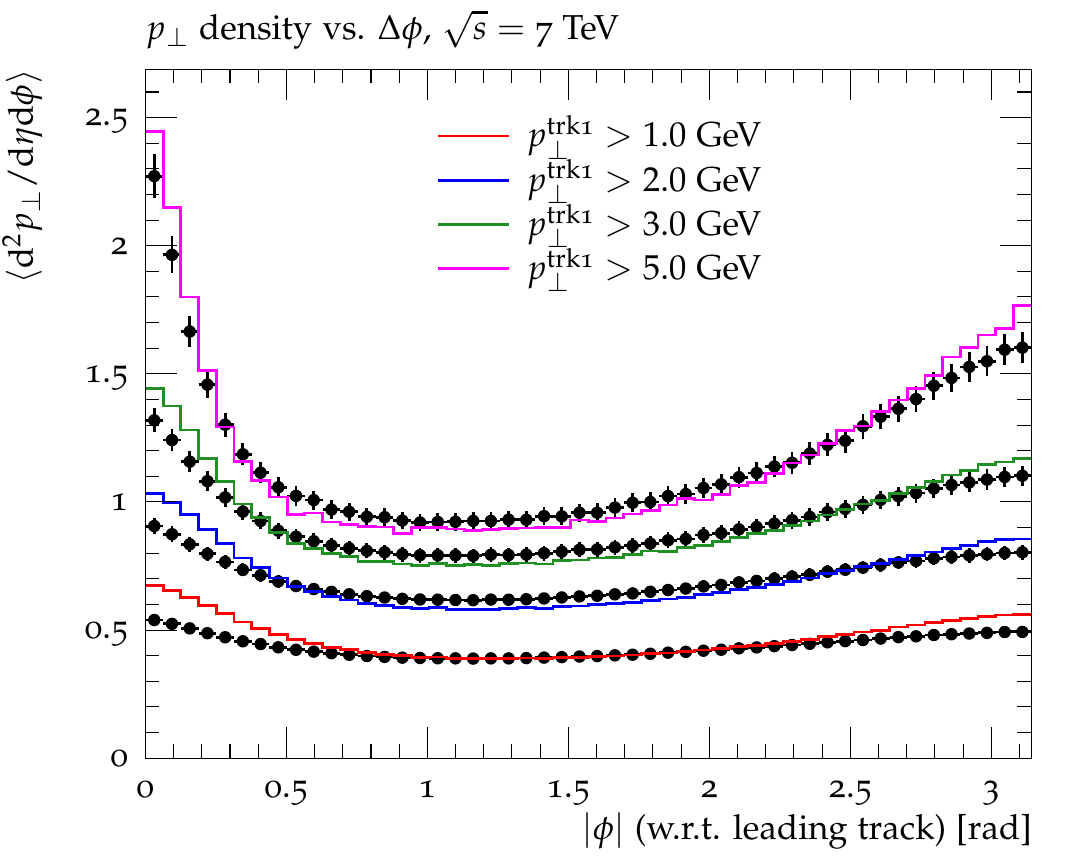}%
    }   
    \caption{Azimuthal distribution of charged particle multiplicity and \ptsum
    densities, with respect to the direction of the leading charged particle (at $\phi =
    0$), for $|\eta| < 2.5$.  The densities are shown for $\ptlead>1$~GeV, $\ptlead>2$~GeV,
    $\ptlead>3$~GeV and $\ptlead>5$~GeV. The data is compared to the UE7-2 tune.}%
    \label{fig:UE7000phi}
  \end{center}
\end{figure}
In the first step we check whether with the new colour reconnection model
allows us to describe the ATLAS MB data at 900 GeV, which was the main
motivation for extending the model.  For that purpose, we tuned the four model
parameters using the Professor tool~\cite{Buckley:2009bj} to the
diffraction-reduced ATLAS MB sample. The results of the tune are shown by the
blue lines in Fig.~\ref{fig:MB900}.

First, from Fig.~\ref{fig:MB900:b} we can see that -- as expected -- colour
reconnection helps to achieve a better description of $\langle p_T\rangle
(N_{\rm ch})$.  Secondly, also the $\eta$ distribution is now well described.
The old MPI model, whose results are comprised in the dashed \HWPP 2.4.2 lines
in Fig.~\ref{fig:MB900}, generates pronounced peaks in the forward directions.
The reason for that behaviour is that the soft scatters in this model are
disconnected from the rest of the event, $p_{\rm disrupt} = 1$. In combination
with the small transverse momenta of the soft scatters, $p_t < \ptmin$, this
colour disruption accounts for a strong population of particles of the forward
region of the event. Changing the value of $p_{\rm disrupt}$ to smaller value,
as done in the \HWPP 2.5 model, helps to get a better shape, however 
 the colour reconnection model is vital to describe this observable. Other MB
observables provided by ATLAS, for instance the charged-particle multiplicity
as a function of the transverse momentum, are also well described with the
extended model.

Finally, for the first time we show the results of the new model against the UE data
collected by ATLAS \cite{Aad:2010fh} at 7 TeV. As before we use the Professor tool to tune the
parameters of the model. This time we used two observables for the tune: The mean number
of stable charged particles per unit of $\eta$-$\phi$, $\dNchgdetadphi$, and the mean
scalar $p_{\perp}$ sum of stable particles per unit of $\eta$-$\phi$, $\dpTsumdetadphi$,
both as a function of $\ptlead$ in the kinematic range $p_{\perp} > 500$~MeV and $|\eta| <
2.5$.
As a result, we obtained a tune named UE7-2, which gives very satisfactory results for the tuned observables.  
The full comparison with all ATLAS UE and MB data sets is available on the \HWPP 
tune page \cite{tune_wiki} -- here we just present a few selected examples. 
In Fig.~\ref{fig:UE7000} we show $\dNchgdetadphi$ and $\dpTsumdetadphi$ as a function 
of $\ptlead$ for the lower $p_{\perp}$ cut ($p_{\perp} > 100$ MeV) in the transverse region 
(which is the most sensitive region with respect to multiple interactions) compared to the \HWPP UE7-2 results.  
The observables  with the lower $p_{\perp}$ cut were not available during the preparation of the tune, 
and these excellent results can therefore be treated as a prediction of the model.
In Fig.~\ref{fig:UE7000sd} we see that the standard deviation of the charged particle
multiplicity and charged particle scalar $\sum p_{\perp}$ densities in the transverse
region, which were not included in the tuning procedure, are also described correctly.  In
the last plot, Fig.~\ref{fig:UE7000phi}, we present the angular distributions of the
charged particle multiplicity and $\sum p_{\perp}$, with respect to the leading charged
particle (at $\Delta \phi=0$). The data sets are shown for four different lower
$p_{\perp}$ cut values in the transverse momentum of the leading charged particle. With
the increase of the leading charged particle $\ptlead$, the development of a jet-like
structure can be observed. The overall description of the data is satisfactory but we can
also see that the description improves as the lower cut value in \ptlead gets higher. 
Finally, the values of the model parameters used in the UE7-2 tune are
$$\ptmin=3.36\ \text{GeV},\ \mu^2 = 0.81\ \text{GeV}^2,\  \pdisrupt = 0.35,\ \preco =
0.616\ .$$

For completeness, having the value of $\mu$,  using Eq.~\ref{eqn:overlap} we can calculate
$\seff=[\int \mathrm{d} \mathbf{b}A^2(b,\mu)]^{-1}=42.28\ \text{mb}$.~Since, as we have shown in 
\cite{tune_wiki}, there is some freedom in choosing the parameters during a tune 
it is possible to describe data at the same
level of accuracy  having  $\mu^2$ in the range between $0.8-1.35\ \text{GeV}^2$.
Therefore, the value of $\seff$ in other tunes can be significantly different to the one calculated above.
In the case of the highest possible value of $\mu^2 = 1.35\ \text{GeV}^2$, we find
$\seff=[\int \mathrm{d} \mathbf{b}A^2(b,\mu)]^{-1}=25.37\ \text{mb}$.




\section{Summary and outlook}
We have shown that introducing colour reconnections and stronger colour
correlations of soft scatters with the beam remnants enables a proper
description of non-diffractive MB observables.  Moreover, we presented a
comparison of the \HWPP UE7-2 tune to ATLAS UE data at 7 TeV.  Despite these
very promising results, there are still open questions concerning the MPI model
in \HWPP, which we would like to address in the future.  In particular, we hope
to obtain a deeper physical understanding of the colour reconnection model.
Also the energy dependence of the model parameters is to be surveyed since,
with the current model, different tunes for different $\sqrt{s}$ are mandatory.
Another physically appealing question is whether and how a united description
of UE and MB data sets can be achieved.

The new model is implemented and available in the current version of \HWPP 2.5.
News concerning \HWPP  tunes are available \cite{tune_wiki}.

\bibliography{Herwig++}
%
%
%
%
%
%
%
%

\end{document}